\documentclass[conference]{IEEEtran}
\IEEEoverridecommandlockouts

\usepackage{amsmath,amssymb,amsfonts}
\usepackage{algorithmic}
\usepackage{graphicx}
\usepackage{textcomp}
\usepackage{xcolor}
\usepackage{subfigure}
\usepackage{braket}
\usepackage{amssymb}
\usepackage{graphicx}
\usepackage{esint}
\usepackage{babel}
\usepackage{color}
\usepackage{wrapfig}

\begin{document}

\title{Solving the Traveling Salesman Problem via Different Quantum Computing Architectures}

\author{\IEEEauthorblockN{Venkat Padmasola}
\IEEEauthorblockA{\textit{Department of Physics} \\
\textit{Center of Quantum Science and Engineering} \\
\textit{Stevens Institute of Technology}\\
Hoboken, N.J., 07030 \\
vpadmaso@stevens.edu}
\and
\IEEEauthorblockN{Zhaotong Li}
\IEEEauthorblockA{\textit{Department of Physics} \\
\textit{Center of Quantum Science and Engineering} \\
\textit{Stevens Institute of Technology}\\
Hoboken, N.J., 07030 \\
Zhli@stevens.edu}
\and
\IEEEauthorblockN{Rupak Chatterjee}
\IEEEauthorblockA{\textit{Department of Applied Physics} \\
\textit{Tandon School of Engineering}\\ 
\textit{New York University} \\
Brooklyn, NY 11201 \\
Rupak.Chatterjee@NYU.edu}
\and
\IEEEauthorblockN{Wesley Dyk}
\IEEEauthorblockA{\textit{Quantum Computing Inc.} \\
\textit{215 Depot Court SE, Suite 215, Leesburg, VA 20175} \\
wdyk@quantumcomputinginc.com}}

\maketitle

\begin{abstract}
We study the application of emerging photonic and quantum computing architectures to solving the Traveling Salesman Problem (TSP), a well-known NP-hard optimization problem. We investigate several approaches: Simulated Annealing (SA), Quadratic Unconstrained Binary Optimization (QUBO-Ising) methods implemented on quantum annealers and Optical Coherent Ising Machines, as well as the Quantum Approximate Optimization Algorithm (QAOA) and the Quantum Phase Estimation (QPE) algorithm on gate-based quantum computers.

QAOA and QPE were tested on the IBM Quantum platform. The QUBO-Ising method was explored using the D-Wave quantum annealer, which operates on superconducting Josephson junctions, and the 
Quantum Computing Inc (QCi) Dirac-1 entropy quantum optimization machine. 
Gate-based quantum computers demonstrated accurate results for small TSP instances in simulation. However, real quantum devices are hindered by noise and limited scalability. Circuit complexity grows with problem size, restricting performance to TSP instances with a maximum of 6 nodes.

In contrast, Ising-based architectures show improved scalability for larger problem sizes. SQUID-based Ising machines can handle TSP instances with up to 12 nodes, while 
entropy computing implemented in hybrid optoelectronic components 
extend this capability to 18 nodes. Nevertheless, the solutions tend to be suboptimal due to hardware limitations and challenges in achieving ground state convergence as the problem size increases. Despite these limitations, Ising machines demonstrate significant time advantages over classical methods, making them a promising candidate for solving larger-scale TSPs efficiently.

\end{abstract}

\section{Introduction}

The Traveling Salesman Problem (TSP) is an NP-hard optimization problem with a wide range of real-world applications, from optimizing delivery routes in supply chains to DNA sequencing \cite{DNAsequencing}. The primary objective of the TSP is to determine a path starting from a given location, visiting every location in a specified list exactly once, and returning to the starting point while minimizing the total distance traveled. This problem involves strict, non-violable constraints, making it challenging to solve efficiently.

While the TSP is conceptually simple and can be solved for small graphs with well-defined edge connectivity, it becomes computationally intractable as the problem size increases. Table~\ref{nphardness} shows the rapid growth in the number of possible paths as the number of nodes increases, highlighting the NP-hard nature of the problem.

\begin{table}[htbp!]
    \centering
    \begin{tabular}{|c|c|}
    \hline
         Number of Nodes & Number of Possible Paths \\
    \hline
        5 & 12\\
    \hline
        10 & 181,000\\
    \hline
        20 & $6.1 \times 10^{16}$\\
    \hline
        25 & $3.1 \times 10^{23}$\\
    \hline
    \end{tabular}
    \caption{Growth in problem size versus number of possible path combinations for the TSP.}
    \label{nphardness}
\end{table}

Significant efforts have been dedicated to the development of efficient algorithms to solve the TSP, given its diverse and impactful applications \cite{BOCK2024, Karlahoffman}. In this paper, we leverage recent advances in computing hardware to explore novel approaches to solving the TSP. Specifically, we focus on quantum simulations, optical/optoelectronic computing, and quantum computing techniques to address the computational challenges posed by this problem.

\section{The TSP Formulation for Gate based Quantum Computing}
\subsection{QAOA Formulation}

\subsubsection{Theoretical formulation for the QAOA}
The quantum approximate optimization algorithm is a hybrid quantum-classical variational algorithm designed to tackle combinatorial optimization problems. An essential  ingredient for understanding and deploying the QAOA is a
constructive approach to carry out the outer-loop classical optimization. The performance of the QAOA on MaxCut problems has revealed its ability to exploit non-adiabatic operations \cite{Lukin_2020}.  For instance, the QAOA method compared to quantum annealing can succeed especially in scenarios adiabatic quantum annealing fail due to a small spectral gap between the ground state and excited states. The comparison reveals that the QAOA can learn via optimization to utilize nonadiabatic mechanisms to circumvent the challenges associated with vanishing spectral gaps \cite{Lukin_2020}. Finally, we
provide a realistic resource analysis on the experimental implementation of the QAOA. When quantum
fluctuations in measurements are accounted for, we illustrate that optimization is important only for
problem sizes beyond numerical simulations but accessible on near-term devices. 

For combinatorial optimization, the quantum approximate optimization algorithm  briefly had a better approximation ratio than any known polynomial time classical algorithm until a more effective classical algorithm was proposed. The relative speed-up of the quantum algorithm is an open research question.

The heart of QAOA relies on the use of unitary operators dependent on $2p$ angles, where $p>1$ is an input integer. These operators are iteratively applied on a state that is an equal weighted quantum superposition of all the possible states in the computational basis. In each iteration, the state is measured in the computational basis and the cost function $f_Q$ is calculated. After a sufficient number of repetitions, the value of $f_Q$  is almost optimal, and the state being measured is close to being optimal as well.

It was also shown that QAOA exhibits a strong dependence on the ratio of a problem's constraint to variables (problem density) \cite{Akshay_2020} placing a limiting restriction on the algorithm's capacity to minimize a corresponding objective function.

It was soon recognized that a generalization of the QAOA process is essentially an alternating application of a continuous-time quantum walk on an underlying graph followed by a quality-dependent phase shift applied to each solution state. This generalized QAOA was termed the QWOA (Quantum Walk-Based Optimization Algorithm) \cite{bennett2021quantumwalkbasedvehiclerouting}.

In the paper \cite{Dalzell2020} the authors conclude that a QAOA circuit with 420 qubits and 500 constraints would require at least one century to be simulated using a classical simulation algorithm running on state-of-the-art supercomputers so that would be sufficient for quantum computational supremacy.

To demonstrate solutions to the TSP using Universal gate based processors, we have explored known formulations of the QAOA, some of which are discussed below with specific processors as the quantum back-ends. Most known ways to formulate the TSP rely on defining the problem as a QUBO problem and then converting to a QAOA \cite{minato2024twostepqaoaenhancingquantum}.

\bigskip

For a given classical "cost" function $ f_C(x):\mathbb {B} ^{n}\rightarrow \mathbb {R}$, an approximate algorithm (for maximization) attempts to find a string $x$ that achieves a desired approximation ratio $r$,
\begin{equation}
    \dfrac{ f_C(x)}{ f_C^{max}} \geq r
\end{equation}
where $f_C^{max}$ is the true maximum.

Consider the operator
\begin{equation}
   \hat{H}_C=f_C(\sigma_z^1, \sigma_z^2, ..., \sigma_z^N) 
\end{equation}
and the Hadamard basis
\begin{equation}
\begin{array}{c}
\mathbf{H} \ket{0} =\left.|+\right\rangle =\dfrac{\left.|0\right\rangle +\left.|1\right\rangle }{\sqrt{2}}=\dfrac{1}{\sqrt{2}}\left[\begin{array}{c}
1\\
1
\end{array}\right] \\\\
\mathbf{H} \ket{1} = \left.|-\right\rangle =\dfrac{\left.|0\right\rangle -\left.|1\right\rangle }{\sqrt{2}}=\dfrac{1}{\sqrt{2}}\left[\begin{array}{c}
1\\
-1
\end{array}\right].
\end{array}
\end{equation} 
Consider an equally weighted superposition of
all $2^{n}$ computational basis states by applying a direct product
of Hadamard operators to $\left.|0\right\rangle $,
\begin{equation}
\mathbf{H^{\otimes n}\left.|0\right\rangle =\mathrm{\dfrac{1}{\sqrt{2^{n}}}\sum_{\mathit{z}=0}^{2^{n}-1}}}\left.|z\right\rangle
\end{equation}
where $\left.|z\right\rangle =|z_{n-1}z_{n-2}z_{n-3}\cdots\,z_{2}z_{1}z_{0}\left.\right\rangle $
represents the bit string corresponding to any integer as $z=2^{0}z_{0}+2^{1}z_{1}+\cdots2^{n-1}z_{n-1}$. Each basis state is a eigenvector of $\hat{H}_C$
\begin{equation}
    \hat{H}_C \ket{z}=f_C(\sigma_z^1, \sigma_z^2, ..., \sigma_z^N)\ket{z} =f_C(z)\ket{z}  
\end{equation}
Let the maximum value $f_C^{max}$, being the largest eigenvalue of $\hat{H}_C$, correspond to the eigenvector $\ket{z_{max}}$ such that
\begin{equation}
\bra{z_{max}} \hat{H}_C \ket{z_{max}}  =  f_C^{max}   
\end{equation}
One cannot in general find the state $\ket{z_{max}}$ but one can search for a state
\begin{equation}
\ket{\psi} = \sum_{z \in \{0,1\}^n} a_z \ket{z}
\end{equation}
which is close as possible to $\ket{z_{max}}$.

\bigskip

\subsubsection{The QAOA algorithm step sequence}
\begin{enumerate}
    \item \textbf{Prepare the Initial State:}  
    Begin with a uniform superposition of all possible computational basis states. For a system of \( n \) qubits, this state is prepared using Hadamard gates applied to each qubit:  
    \begin{equation}
        \ket{\psi_0} = 
\mathbf{H^{\otimes n}\left.|0\right\rangle =\mathrm{\dfrac{1}{\sqrt{2^{n}}}\sum_{\mathit{z}=0}^{2^{n}-1}}}\left.|z\right\rangle
    \end{equation}

    \item \textbf{Construct the Problem Hamiltonian:}  
    Define a Hamiltonian \( H_C \) that encodes the optimization problem's objective function. For example, if the goal is to minimize a cost function \( C(z) \), then:
    \begin{equation}
   \hat{H}_C \ket{z}=f_C(\sigma_z^1, \sigma_z^2, ..., \sigma_z^N)\ket{z} =f_C(z)\ket{z}  
    \end{equation}
    where \( f_C(z) \) is the cost associated with the computational basis state \( \ket{z} \).

    \item \textbf{Construct the Mixer Hamiltonian:}  
    Define a mixer Hamiltonian \( H_B \) that facilitates transitions between basis states. The typical choice is:
    \begin{equation}
        \hat{H}_B = \sum_{i=1}^n X_i,
    \end{equation}
    where \( X_i \) is the Pauli-X operator acting on the \( i \)-th qubit. This Hamiltonian helps explore the solution space by driving transitions between states.

    \item \textbf{Exponentiate and Parameterize:}  
    Apply the problem Hamiltonian and mixer Hamiltonian alternately in \( p \) layers (referred to as the '$p$-level' algorithm). Each layer is parameterized by angles \( \gamma \) and \( \beta \), which control the evolution times. The resulting state after \( p \) steps is:
    \begin{equation}
        \ket{\psi_p(\vec{\gamma}, \vec{\beta})} = \prod_{j=1}^p e^{-i \beta_j \hat{H}_B} e^{-i \gamma_j \hat{H}_C} \ket{\psi_0}.
    \end{equation}

    \item \textbf{Optimize Parameters:}  
    Use a classical optimization algorithm to adjust the \( 2p \) parameters \( \vec{\gamma} = (\gamma_1, \dots, \gamma_p) \) and \( \vec{\beta} = (\beta_1, \dots, \beta_p) \) to maximize the expectation value of the problem Hamiltonian:
    \begin{equation}
        \langle H_C \rangle = \bra{\psi_p(\vec{\gamma}, \vec{\beta})} H_C \ket{\psi_p(\vec{\gamma}, \vec{\beta})}.
    \end{equation}
    The classical optimizer iteratively improves these parameters to find the optimal solution.
\end{enumerate}

\subsubsection{Simulation Results on the IBMQ Platform for the QAOA Algorithm}

\begin{figure}[htbp!]
\centering
\includegraphics[width=1.0\linewidth]{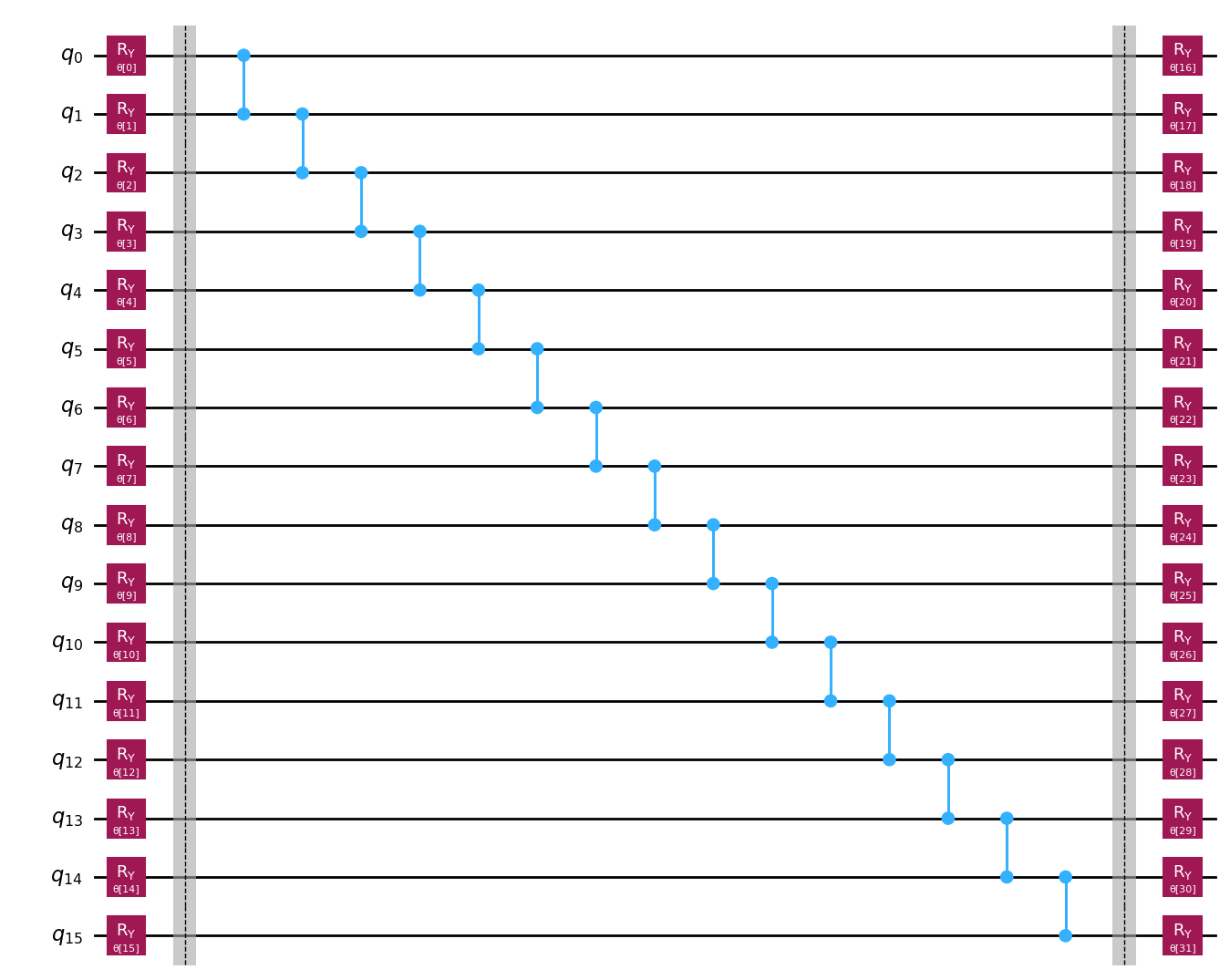}
\caption{A single layer of the six-layer QAOA quantum circuit.}
\label{fig:QAOA_circuit}
\end{figure}

We initialize an arbitrary set of angles between \( 0 \) and \( 2\pi \), which are passed through a classical COBYLA optimizer to determine the optimal set of angles. For a 4-node TSP, a 16-qubit circuit is initialized, as the qubit requirement scales as \( N^2 \). The values pass through a quantum circuit with a depth of 6 layers. Therefore, there are 96 angles serving as hyperparameters for optimization. The output is then measured. Figure~\ref{fig:QAOA_circuit} illustrates a single layer of the six-layer QAOA quantum circuit, where the hyperparameters are encoded into the \( R_Y \) rotation gates.

\begin{figure}[htbp!]
\centering
\includegraphics[width=1.0\linewidth]{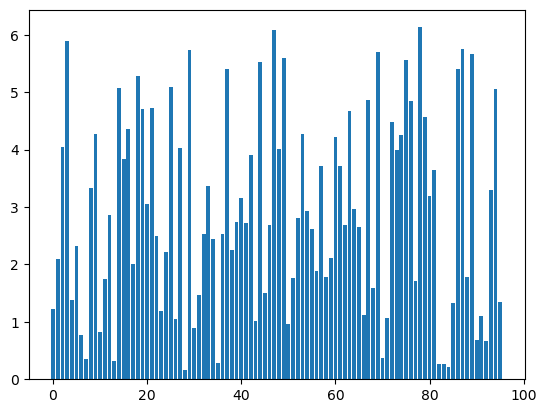}
\caption{Initial set of angles before optimization.}
\label{fig:before_optimization}
\end{figure}

\begin{figure}[htbp!]
\centering
\includegraphics[width=1.0\linewidth]{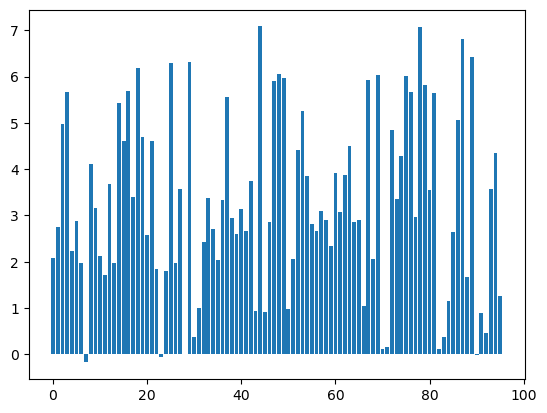}
\caption{Final set of angles obtained after optimization.}
\label{fig:after_optimization}
\end{figure}

The bar plots for the 96 angles before and after optimization are presented in Figures~\ref{fig:before_optimization} and \ref{fig:after_optimization}, respectively.

\begin{figure}[htbp!]
\centering
\includegraphics[width=1.0\linewidth]{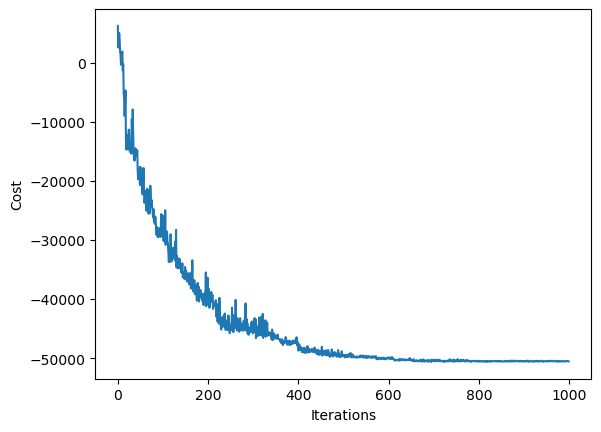}
\caption{Convergence to the optimal set of angles to determine the ground state of the QAOA Hamiltonian.}
\label{fig:QAOA_convergence}
\end{figure}

Figure~\ref{fig:QAOA_convergence} shows the convergence to the minimum cost during the optimization process using the QAOA approach.

For a Hamiltonian of size \( 16 \times 16 \), there are \( 2^{16} \) basis states. The measurement outcomes clearly demonstrate the need for an optimized set of angles for the \( R_Y \) rotation gates. A classical optimization step is required to fine-tune these angles such that the Hamiltonian returns the correct solution with maximum probability. Fine-tuning the \( R_Y \) angle values is absolutely essential to ensure accurate solutions. Without this optimization, the results may violate constraints and yield incorrect solutions.

This observation highlights a significant limitation of the QAOA approach: the classical optimization step introduces a time bottleneck. As a result, QAOA cannot ensure polynomial time complexity due to the reliance on classical hyperparameter optimization.

\subsection{The QPE Formulation}
\subsubsection{Overview of the Quantum Phase Estimation Algorithm}
The Quantum Phase Estimation (QPE) algorithm extracts the phase \( \theta \) associated with an eigenvalue of a unitary operator \( U \) \cite{srinivasan2018efficient}. Specifically, if \( U |\psi\rangle = e^{2\pi i \theta} |\psi\rangle \), QPE estimates the phase \( \theta \) with \( n \)-bit precision. The QPE circuit operates on two registers: the \textbf{upper register} of \( n \) qubits, initialized in \( |0\rangle^{\otimes n} \), is used for phase estimation, and the \textbf{lower register} is initialized in the eigenstate \( |\psi\rangle \) of \( U \), such that \( U |\psi\rangle = e^{2\pi i \theta} |\psi\rangle \). The key steps of QPE are as follows: 

1) Apply Hadamard gates on all qubits of the upper register to create a superposition state. 

2) Apply controlled-\( U^{2^j} \) operations, where \( j \) corresponds to each qubit in the upper register. 

3) Apply the inverse Quantum Fourier Transform (QFT) on the upper register to extract the phase \( \theta \). The final state of the upper register encodes the binary representation of \( \theta \).

\subsubsection{Constructing the Unitary Operator \( U \)}
The unitary matrix \( U \) operates on a Hilbert space of dimension \( 2^m \), where \( m \) is the number of qubits in the lower register. The action of \( U \) on an eigenstate \( |\psi\rangle \) is:
\begin{equation}
U |\psi\rangle = e^{2\pi i \theta} |\psi\rangle.
\end{equation}
The QPE algorithm requires the controlled-\( U \) operator, defined as:
\begin{equation}
C\text{-}U = |0\rangle\langle 0| \otimes I + |1\rangle\langle 1| \otimes U.
\end{equation}
Here, \( |0\rangle \) and \( |1\rangle \) correspond to the control qubit states, and \( I \) is the identity matrix. The controlled-\( U^{2^j} \) operator generalizes this to:
\begin{equation}
C\text{-}U^{2^j} = |0\rangle\langle 0| \otimes I + |1\rangle\langle 1| \otimes U^{2^j}.
\end{equation}

\subsubsection{Phase Kickback and Superposition State}
The Hadamard gate applied to the \( n \)-qubit upper register produces a superposition state:
\begin{equation}
|0\rangle^{\otimes n} \xrightarrow{H^{\otimes n}} \frac{1}{2^{n/2}} \sum_{k=0}^{2^n-1} |k\rangle,
\end{equation}
where \( |k\rangle \) represents the computational basis states. The controlled-\( U^{2^j} \) gates apply phase shifts based on the eigenvalue \( e^{2\pi i \theta} \). For the \( j \)-th control qubit, the operation introduces a phase factor \( e^{2\pi i \cdot 2^j \theta} \), resulting in the following state:
\begin{equation}
\frac{1}{2^{n/2}} \sum_{k=0}^{2^n-1} |k\rangle \otimes U^{k} |\psi\rangle.
\end{equation}
By leveraging the eigenvalue relationship \( U |\psi\rangle = e^{2\pi i \theta} |\psi\rangle \), the lower register accumulates phase factors:
\begin{equation}
\frac{1}{2^{n/2}} \sum_{k=0}^{2^n-1} e^{2\pi i k \theta} |k\rangle \otimes |\psi\rangle.
\end{equation}
The upper register now contains the phase information encoded in the amplitudes of the basis states \( |k\rangle \).

\subsubsection{Applying the Inverse QFT}
To extract the phase \( \theta \), we apply the inverse Quantum Fourier Transform (QFT) on the upper register. The QFT maps the amplitudes \( e^{2\pi i k \theta} \) into the binary representation of \( \theta \) with high probability. The inverse QFT is defined as:
\begin{equation}
\text{QFT}^{\dagger} |k\rangle = \frac{1}{2^{n/2}} \sum_{j=0}^{2^n-1} e^{-2\pi i k j / 2^n} |j\rangle.
\end{equation}
After applying \( \text{QFT}^{\dagger} \), the upper register collapses to a computational basis state \( |j\rangle \), where \( j \) represents the binary approximation of \( 2^n \theta \). That is, the output of the QPE algorithm is the \( n \)-bit approximation of the phase \( \theta \), such that:
\begin{equation}
\theta \approx \frac{j}{2^n}, \quad j \in \{0, 1, \dots, 2^n-1\}.
\end{equation}

\subsubsection{Encoding Hamiltonian Cycles}
If QPE is applied to a system where the unitary \( U \) encodes Hamiltonian cycles as needed for the TSP problem, the eigenstates \( |\psi\rangle \) are expressed as:
\begin{equation}
|\psi\rangle = \bigotimes_{j=1}^n |i(j)-1\rangle_2,
\end{equation}
where \( i(j) \) represents the index of the city visited at step \( j \). For example, for the sequence \( 1-2-4-3 \):
\begin{align}
|i(1)-1\rangle &= |2\rangle = |10\rangle_2, \\
|i(2)-1\rangle &= |0\rangle = |00\rangle_2, \\
|i(4)-1\rangle &= |1\rangle = |01\rangle_2, \\
|i(3)-1\rangle &= |3\rangle = |11\rangle_2.
\end{align}
The complete encoded state is:
\begin{equation}
|\psi\rangle = |10\rangle \otimes |00\rangle \otimes |01\rangle \otimes |11\rangle = |10000111\rangle.
\end{equation}

\bigskip

\begin{figure}[htbp!]
\centering
\includegraphics[width=1.0\linewidth]{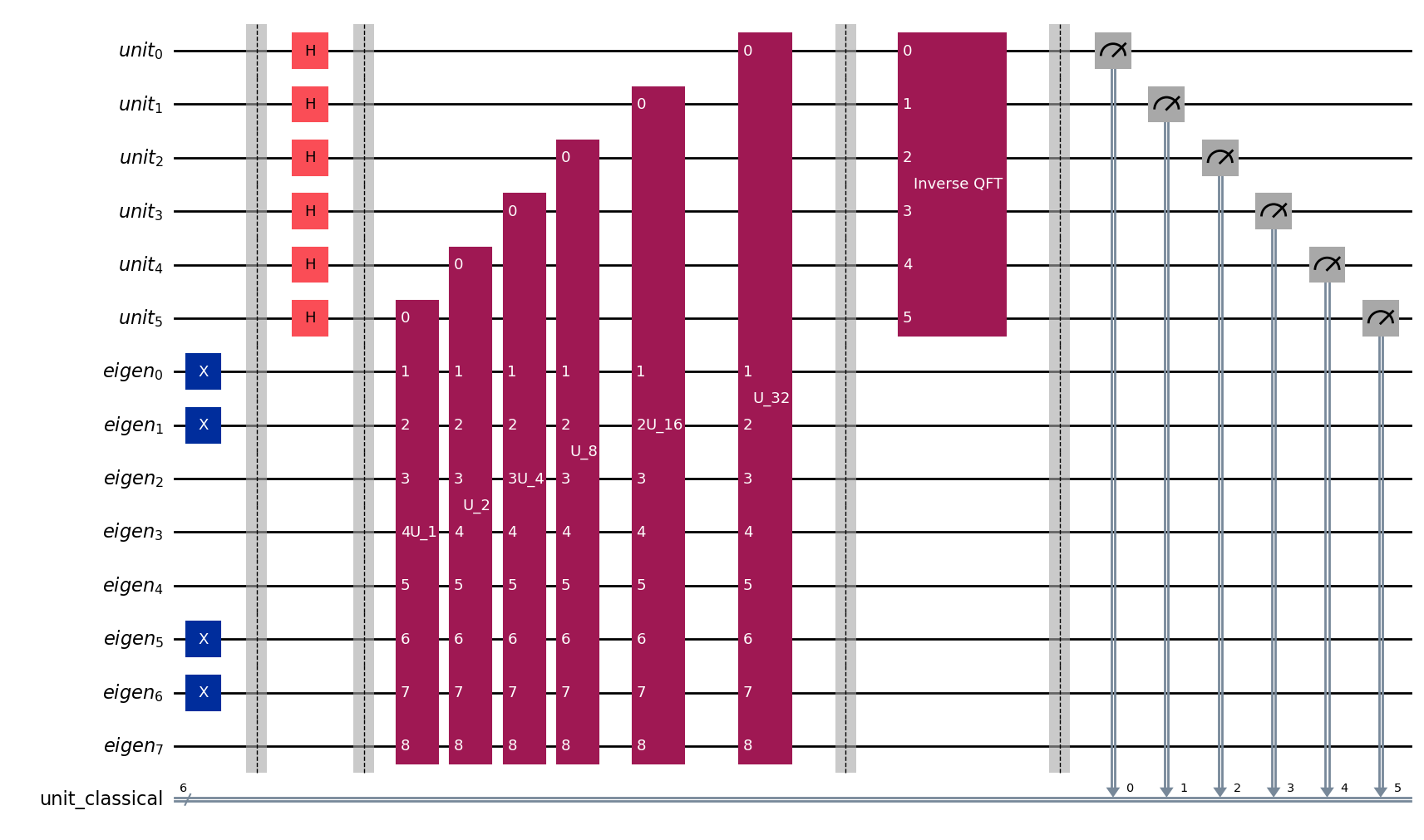}
\caption{Circuit for Quantum Phase Estimation(QPE)}
\label{QPE_circuit}
\end{figure}

In its simplest form, the quantum phase estimation (QPE) circuit operates on two registers of qubits. The lower register is repeatedly acted on by a controlled-\( U \) operator, where the control qubits are in the upper register. An inverse quantum Fourier transform (QFT) is then applied to the upper register, which serves as the estimator for an \( n \)-bit estimate of the phase \( \theta \). Applying this circuit block to a 16-qubit case, we obtain the following quantum circuit. Figure~\ref{QPE_circuit} illustrates the basic setup for QPE.

To encode any arbitrary distance matrix to a format suitable for the QPE algorithm, we first normalize the values in raw distance matrix such that all the values are in between 0 and 1. We do this so that any such value can be represented by euler phase angles(the exponents are the phase angles which lie between $0$ and $2 \pi$ ).

The matrix $B$ is a representation of the normalized distance matrix encodes as euler phases.
\begin{equation}
B=\left[\begin{array}{llll}
e^{i \phi_{1 \rightarrow 1}} & e^{i \phi_{1 \rightarrow 2}} & e^{i \phi_{1 \rightarrow 3}} & e^{i \phi_{1 \rightarrow 4}} \\\\
e^{i \phi_{2 \rightarrow 1}} & e^{i \phi_{2 \rightarrow 2}} & e^{i \phi_{2 \rightarrow 3}} & e^{i \phi_{2 \rightarrow 4}} \\\\
e^{i \phi_{3 \rightarrow 1}} & e^{i \phi_{3 \rightarrow 2}} & e^{i \phi_{3 \rightarrow 3}} & e^{i \phi_{3 \rightarrow 4}} \\\\
e^{i \phi_{4 \rightarrow 1}} & e^{i \phi_{4 \rightarrow 2}} & e^{i \phi_{4 \rightarrow 3}} & e^{i \phi_{4 \rightarrow 4}}
\end{array}\right]
\end{equation}

Since we have a 4 node distance matrix, we then construct 4 unitary operators \( U_j \) from \( B \) such that \((U_j)_{k,k} = [B]_{j,k}\): 

\begin{equation}
U_j = \text{diag}\left(B_{j,1}, B_{j,2}, B_{j,3}, B_{j,4}\right)
\end{equation}

Finally, the tensor product of the 4 \( U_j \) operators \( U = U_1 \otimes U_2 \otimes U_3 \otimes U_4 \) will also be a unitary matrix.

We define $U_{j}$ generally as
\begin{equation}
U_{j}=\left(\sum_{k=1}^{n} B[j][k] \times \text { outer product of basis vectors }\right) 
\end{equation}
That is,
$$
\begin{aligned}
& U_{1}=\left[\begin{array}{cccc}
e^{i \phi_{1 \rightarrow 1}} & 0 & 0 & 0 \\
0 & e^{i \phi_{2 \rightarrow 1}} & 0 & 0 \\
0 & 0 & e^{i \phi_{3 \rightarrow 1}} & 0 \\
0 & 0 & 0 & e^{i \phi_{4 \rightarrow 1}}
\end{array}\right] \\
& U_{2}=\left[\begin{array}{cccc}
e^{i \phi_{1 \rightarrow 2}} & 0 & 0 & 0 \\
0 & e^{i \phi_{2 \rightarrow 2}} & 0 & 0 \\
0 & 0 & e^{i \phi_{3 \rightarrow 2}} & 0 \\
0 & 0 & 0 & e^{i \phi_{4 \rightarrow 2}}
\end{array}\right] \\
& U_{3}=\left[\begin{array}{cccc}
e^{i \phi_{1 \rightarrow 3}} & 0 & 0 & 0 \\
0 & e^{i \phi_{2 \rightarrow 3}} & 0 & 0 \\
0 & 0 & e^{i \phi_{3 \rightarrow 3}} & 0 \\
0 & 0 & 0 & e^{i \phi_{4 \rightarrow 3}}
\end{array}\right] \\
& U_{4}=\left[\begin{array}{cccc}
e^{i \phi_{1 \rightarrow 4}} & 0 & 0 & 0 \\
0 & e^{i \phi_{2 \rightarrow 4}} & 0 & 0 \\
0 & 0 & e^{i \phi_{3 \rightarrow 4}} & 0 \\
0 & 0 & 0 & e^{i \phi_{4 \rightarrow 4}}
\end{array}\right]
\end{aligned}
$$
Creating the final unitary matrix $U$ from the series of submatrices $U_{i}$ The way $U$ defined is:
$$
U=U_{1} \otimes U_{2} \otimes U_{3} \otimes U_{4}
$$

\subsubsection{Simulation and experimental results for the QPE algorithm on IBMQ platform}

The QPE method has certain advantages over the QAOA approach significantly due to the fact that for a problem of size n, the qubit requirement in the case of QPE being $n\log n$ as opposed to the $n^2$ qubit requirement in the case of QAOA.

\begin{figure}[htbp!]
\centering
\includegraphics[width=1.0\linewidth]{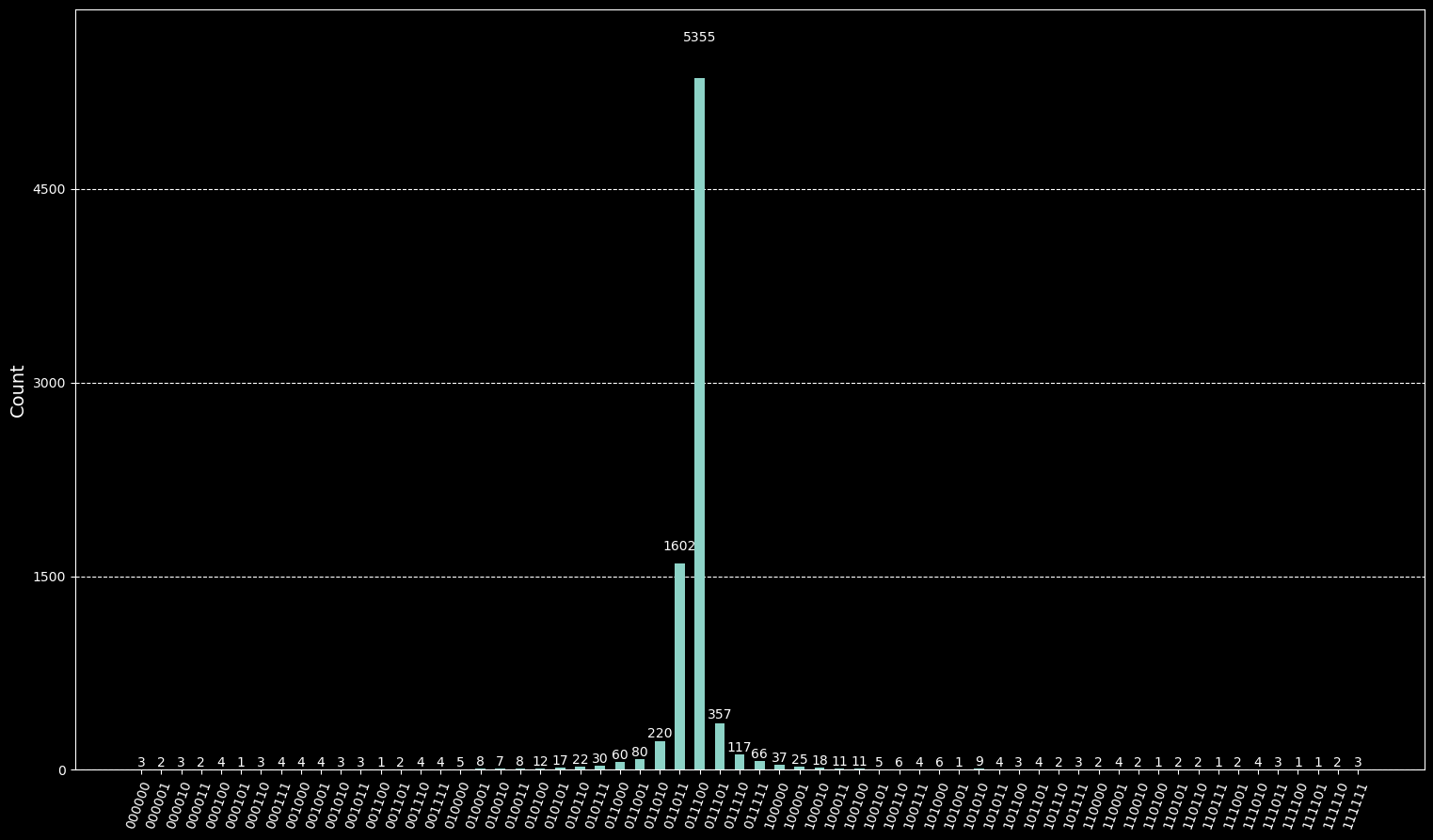}
\caption{Histogram of the measurement results for 8192 shots for the QPE using the IBM QASM simulator}
\label{"Histogram of the measurement QASM simulator"}
\end{figure}

\begin{figure}[htbp!]
\centering
\includegraphics[width=1.0\linewidth]{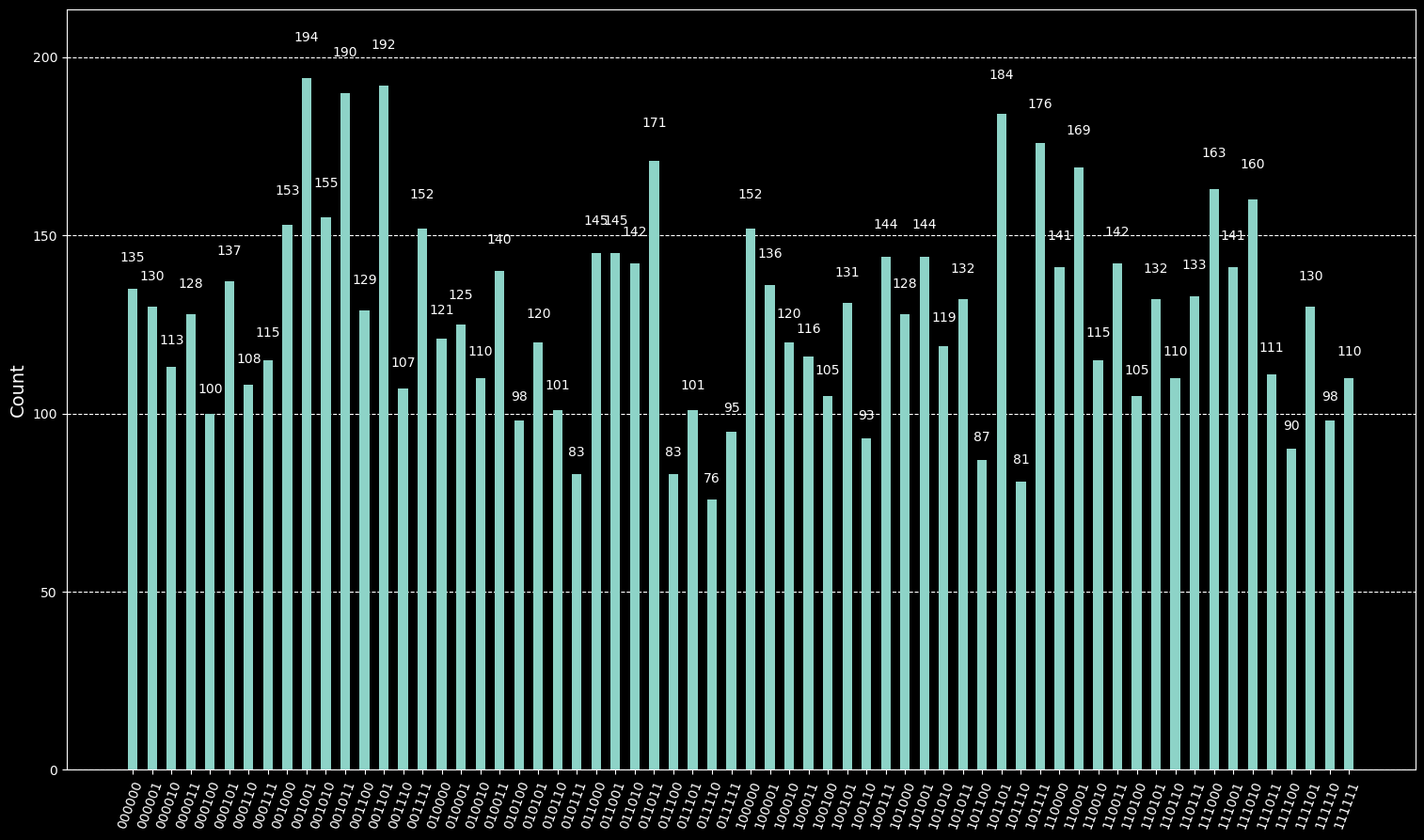}
\caption{Histogram of the measurement results for 8192 shots for the QPE using the IBM Brisbane Quantum Computer}
\label{"Histogram IBM Brisbane Quantum Computer"}
\end{figure}

It is evident that real quantum computers are prone to errors due to noise, as illustrated in Figure \ref{"Histogram IBM Brisbane Quantum Computer"}, while the simulation results yield more accurate results, as shown in Figure \ref{"Histogram of the measurement QASM simulator"}. Previous studies have demonstrated that NISQ gate-based quantum computers are not yet sufficiently advanced for fault tolerance or large enough to achieve quantum advantage \cite{noisyNISQ}. To obtain reliable outputs from universal gate-based quantum computers in the NISQ era, quantum error correction or error mitigation techniques are essential \cite{NISQerrormitigation}. Consequently, it is imperative to explore alternative computing architectures that can potentially provide polynomial-time advantages. In the following sections, we investigate alternative approaches to solving the TSP using Ising-type simulators and hardware.

\section{The TSP Formulation for Ising type simulators and hardware}
\subsection{Ising and QUBO Models}
The Ising model and its equivalent Quadratic Unconstrained Binary Optimization (QUBO) formulation constitute a central problem class for adiabatic quantum computation (AQC), where they are solved through a physical process known as quantum annealing.

In this model, we consider a set $\Lambda$ of lattice sites, each with a set of adjacent sites forming a lattice or graph structure. For each lattice site $k \in \Lambda$, there is a discrete variable $\sigma_k$ such that $\sigma_k \in {+1, -1}$, representing the spin states of a system. Any two adjacent sites, or ``nearest neighbors,'' $i, j \in \Lambda$ interact with an exchange coefficient $J_{ij}$. Additionally, each site $j \in \Lambda$ may experience an external magnetic field $h_j$. The energy of a spin configuration $\sigma = (\sigma_1, \sigma_2, ...)$ is given by the following Hamiltonian:

\begin{equation}
H(\sigma ) = -\sum_{\langle i, j \rangle }J_{ij}\sigma _{i}\sigma _{j} - \mu \sum_{j} h_{j}\sigma_{j},
\label{Ising_Ham}
\end{equation}
where $\mu$ represents the magnetic moment.

A specific spin configuration $\sigma$ follows a probability distribution given by the Boltzmann distribution with inverse temperature $\beta \geq 0$:
\begin{equation}
P_{\beta }(\sigma ) = \frac{e^{-\beta H(\sigma )}}{Z_{\beta }},
\end{equation}
where the partition function $Z_{\beta }$ is defined as:
\begin{equation}
Z_{\beta } = \sum _{\sigma } e^{-\beta H(\sigma )}.
\end{equation}
The expectation value of a function $f$ over spin configurations is given by:
\begin{equation}
\langle f \rangle _{\beta } = \sum_{\sigma } f(\sigma ) P_{\beta }(\sigma ).
\end{equation}

The QUBO model is an alternative formulation of the Ising model that replaces spin variables ($\pm1$) with binary variables ($0,1$). This model is widely used in combinatorial optimization, including applications in finance, machine learning, and logistics. The QUBO formulation is NP-hard and often serves as a bridge between classical and quantum approaches to optimization \cite{Stogiannos2022}. For many classical problems such as the maximum cut graph problem, embeddings into a QUBO formulation are well known \cite{Huang2022}. Embeddings for machine learning models include support vector machines \cite{clustering}, clustering, and probabilistic graphical models \cite{QUBOforML}. An almost equivalent physical model,  the  Ising model, has a well known QUBO formulation.

The transformation from the Ising model to QUBO is straightforward and is achieved using the following linear transformation:
\begin{equation}
x_i = \frac{1 + \sigma _{i}}{2}, \quad x_j = \frac{1 + \sigma _{j}}{2}.
\end{equation}
This ensures that for any $\sigma_i, \sigma_j \in {+1, -1}$, we obtain binary variables $x_i, x_j \in {0, 1}$.

Consider a quadratic function $f: \mathbb{B}^{n} \to \mathbb{R}$, defined over an $n$-dimensional binary space:
\begin{equation}
f_{Q}(x) = \sum_{i=1}^{n} \sum _{j=1}^{i} q_{ij} x_{i} x_{j},
\end{equation}
where $q_{ij} \in \mathbb{R}, 1 \leq j \leq i \leq n$, and $x_{i} \in \mathbb{B} = {0,1}$. The goal of QUBO is to find an $n$-dimensional binary vector $x^{}$ that minimizes this function:
\begin{equation}
x^{} = \arg \min_{x \in \mathbb{B}^{n}} f_{Q}(x).
\end{equation}
The QUBO function $f_{Q}$ can be rewritten using a symmetric QUBO matrix $Q \in \mathbb{R}^{n \times n}, Q = [q_{ij}]$:
\begin{equation}
f_{Q}(x) = x^{\top} Q x.
\end{equation}
This matrix representation allows QUBO problems to be efficiently mapped onto quantum annealers or specialized quantum hardware for solving large-scale optimization problems.

\subsection{The TSP Formulation for Simulated Annealing}
Simulated Annealing (SA) is a classical optimization algorithm inspired by the annealing process in metallurgy, where a material is gradually cooled to remove defects and reach a stable low-energy state. SA applies this concept to optimization problems by slowly decreasing the system's temperature, allowing it to explore the solution space and settle into a low-energy configuration. However, SA can be computationally expensive due to its reliance on sequential updates of spin states. Nonetheless, it provides a unique approach to solving the Traveling Salesman Problem (TSP) by leveraging principles that differ fundamentally from universal gate-based quantum machines.

Mathematically, SA closely resembles the formulation of an Ising-type quantum computer. It explores the entire solution space, enabling it to escape local optima. Unlike gradient-based methods, SA does not get trapped in suboptimal solutions since it incorporates randomness to explore neighboring states. This stochastic nature helps the algorithm bypass local minima. However, in highly complex landscapes with numerous local minima, especially for large problem sizes, SA may fail to reach the true ground state solution.

The annealing process involves gradually lowering the "temperature," allowing the algorithm to initially explore a broad set of solutions, including suboptimal ones, before progressively converging toward an optimal or near-optimal solution. This approach enables SA to handle complex, non-convex objective functions without requiring derivatives or assumptions about the function’s structure. It performs effectively even in noisy or irregular problem spaces.

Despite its advantages, SA is not always the fastest or most precise optimization method. Depending on the problem, alternative approaches such as genetic algorithms, particle swarm optimization, or gradient-based techniques may yield better results. Additionally, selecting appropriate coefficients and tuning parameters for SA is critical, as improper settings can significantly affect its performance. A systematic strategy for parameter optimization is necessary to enhance the algorithm’s efficiency and effectiveness in solving large-scale optimization problems.

\subsubsection{Hamiltonian formulation for Simulated Annealing}
\begin{figure}[htbp!]
\centering
\includegraphics[width=0.9\linewidth]{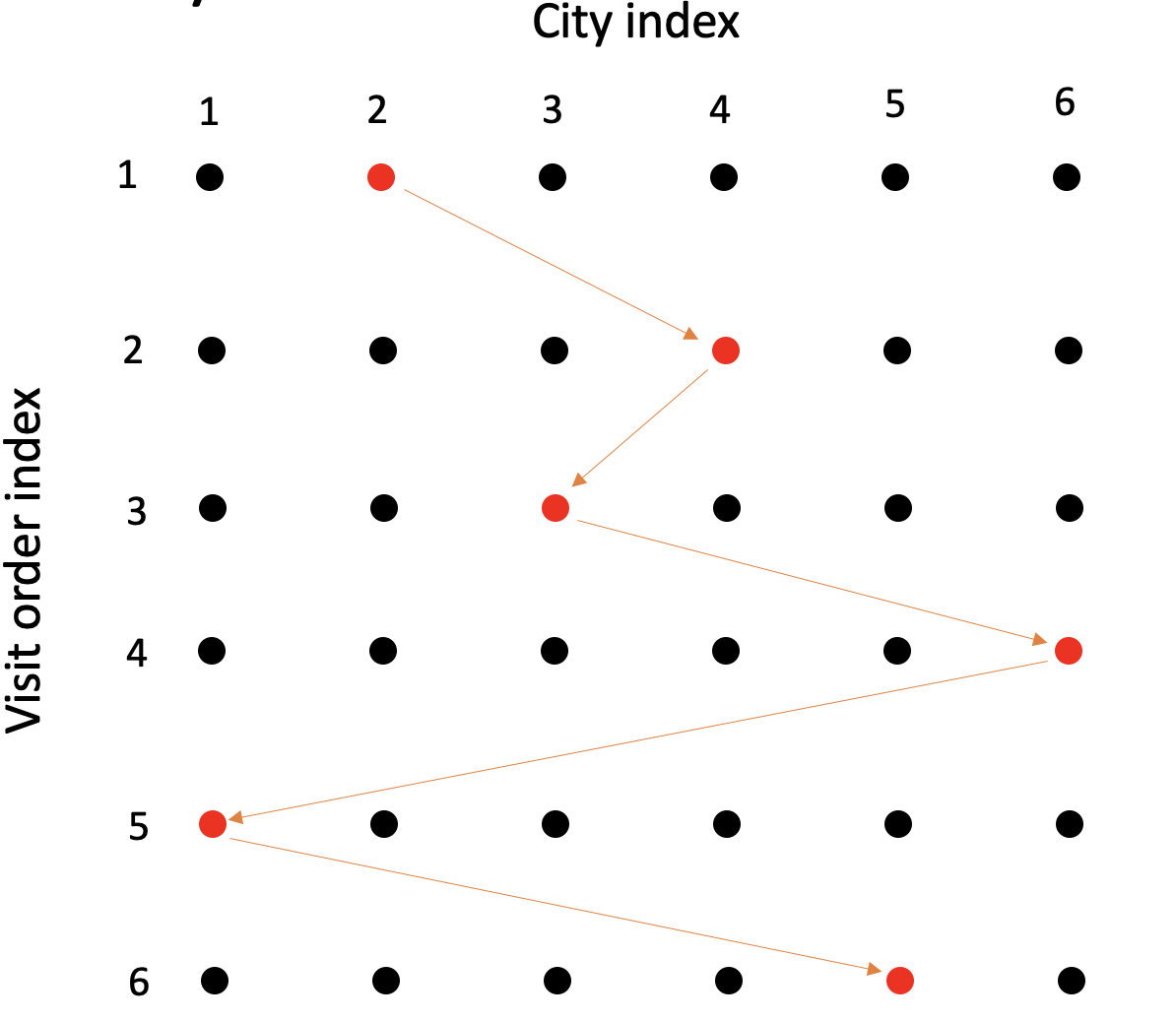}
\caption{routing for a TSP solution under QUBO formulation. The spins are arranged in two dimensions, where the x-axis represents the index of each city and the y-axis is the visitation order}
\label{"Spins mapped to an SLM"}
\end{figure}

One of the most straightforward ways to solve the TSP is to convert it into a Quadratic Unconstrained Binary Optimization (QUBO) problem. For an unconstrained TSP problem, we can use a binary variable $a_{ik}$ to represent whether the city k is visited ($a_{ik}=1$) or not ($a_{ik}=0$) at the i-th step, shown in figure \ref{"Spins mapped to an SLM"}. The objective function can therefore be defined in a quadratic form:
\begin{equation}
H_{obj} = \sum_{k\neq l} \sum_i W_{kl} a_{ik} a_{(i+1)l}
\end{equation}

Where $W_{kl}$ is the element of the distance matrix for distance between city k and city l. Since each city should only be visited once, and only a single city can be visited simultaneously, we introduce the two constrains:
\begin{equation}
H_{cons} = \sum_i(\sum_k a_{ik} - 1)^2 + \sum_k(\sum_i a_{ik} - 1)^2
\end{equation}

If we replace $a_{ik} = (\sigma_{ik}+1) / 2$ where $\sigma_{ik} = \pm 1$ represents an Ising spin, the QUBO formulation is mathematically equivalent to an Ising Hamiltonian:
\begin{eqnarray}
H &=& H_{obj} + \gamma H_{cons} \\ \notag
&=& \frac{1}{4}\sum_{k\neq l} \sum_i W_{kl} \sigma_{ik}\sigma_{(i+1)l} + \frac{1}{2}\sum_{k\neq l}\sum_i W_{kl}\sigma_{ik} \\ \notag
&\quad& + \frac{\gamma}{4}\sum_i \sum_k \sum_l \sigma_{ik}\sigma_{il} + \frac{\gamma}{2}(n-2)\sum_i \sum_k \sigma_{ik} \\ \notag
&\quad& + \frac{\gamma}{4}\sum_k \sum_i \sum_j \sigma_{ik}\sigma_{jk} + \frac{\gamma}{2}(n-2)\sum_k \sum_i \sigma_{ik} \\ \notag
&\quad& + const.
\label{"SA Hamiltonian"}
\end{eqnarray}

\begin{figure}[htbp!]
\centering
\includegraphics[width=0.9\linewidth]{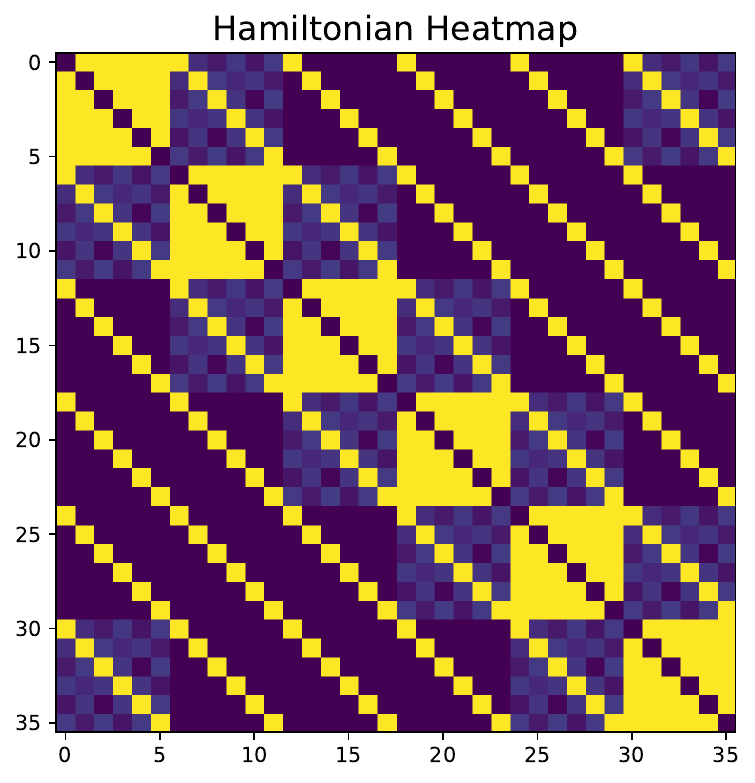}
\caption{Hamiltonian matrix for 6 cities}
\label{"Heatmap of the hamiltonian matrix"}
\end{figure}

For a randomly generated distance matrix W, figure ~\ref{"Heatmap of the hamiltonian matrix"} shows the interaction matrix J reshaped into 2D when $\gamma=5$. 

To demonstrate how the Hamiltonian appears after the adjacency matrix transforms under the equations described above, a heat map of the Hamiltonian for a 6 city TSP is plotted as shown in figure \ref{"Heatmap of the hamiltonian matrix"}.

\subsubsection{Solution for a 6 city TSP using Simulated Annealing}
\begin{figure}[htbp!]
\centering
\includegraphics[width=0.9\linewidth]{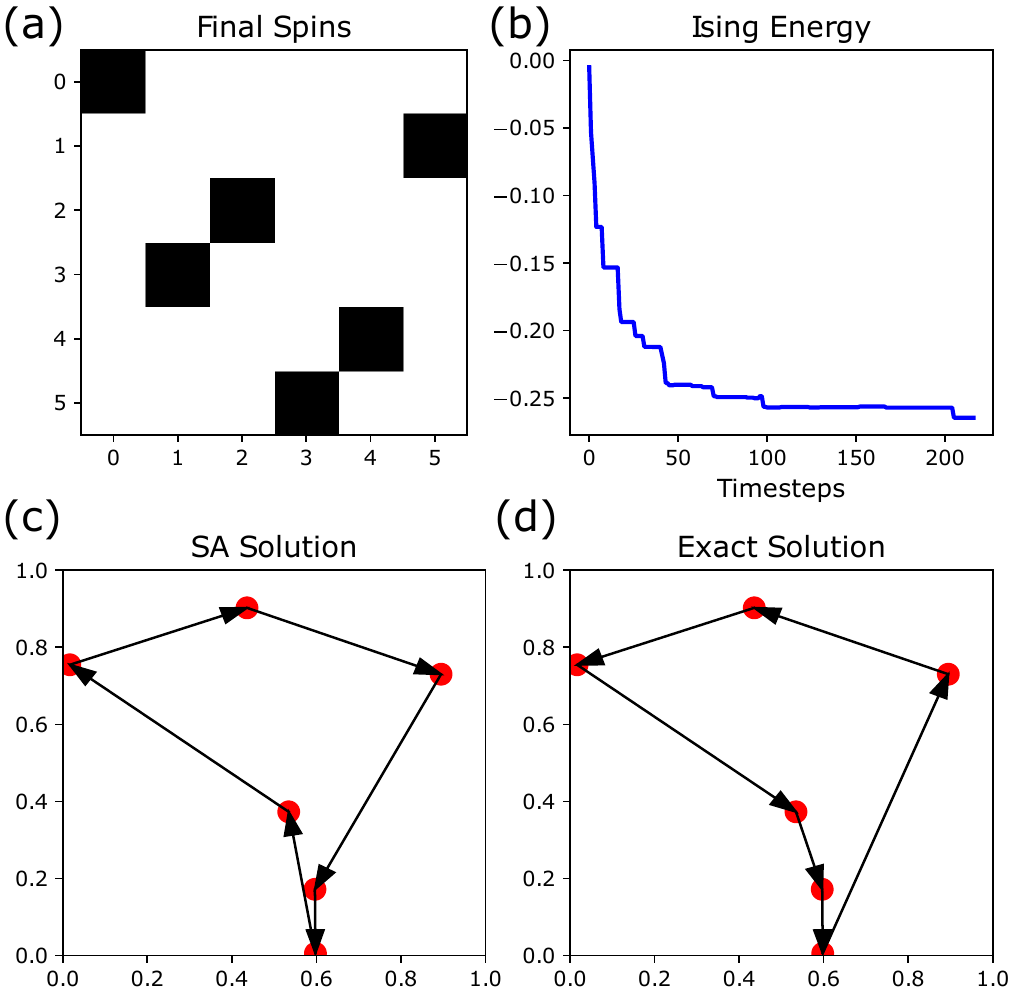}
\caption{Simulated annealing solution for a 6 city TSP. (a) The solution of a 6-city TSP by SA. (b) The Ising energy as a function of the number of time steps. (c) The directed graph of the solution obtained by SA. (d) The directed graph of the solution obtained by the brute force method.}
\label{"Simulated annealing solution and convergence"}
\end{figure}

The solution obtained from this Hamiltonian is initially represented as a linear array of 36 spins, which is then reshaped into a $6 \times 6$ square matrix encoding the solution path, ensuring that each row and column contains exactly one +1 spin value. Figure \ref{"Simulated annealing solution and convergence"} illustrates the plot of Ising energy converging to the lowest value over the time steps,  demonstrating the working mechanism of the simulated annealing algorithm. The final low-energy solution is visualized both as a heat map and a directed graph.

\subsection{Experimental Observations for Adiabatic Quantum computers}
\subsubsection{Hamiltonian formulation for the D-Wave platforms}

Consider $N$ ``cities'' (vertices in a graph) indexed as $\{0,1,...,N-1\}$. Consider the case of a round trip returning to the original city (Hamilton cycle). As each city maybe visited only once, there will be $N$ time steps indexed as $\{1,...,N\}$ where we start at $t=0$. Consider $N+1$ sets of $N+1$ binary variables
\begin{equation}
x_{i,t} \in \mathbb{B}, \;\; i\in \{0,1,...,N\} , \;\; t\in \{0,1,...,N\}
\end{equation}
$x_{i,t}=1$ denotes that the $i^{th}$ city was reached in the $t^{th}$ time. Let us start at the $0^{th}$ city such that we have a boundary condition of $x_{0,0}=1$. The last city visited at $t=N$ must be the original city, so we have another boundary condition $x_{N,N}=1$ where city $N$ is required to be equal to city $0$ (recall that the unique cities are indexed as $\{0,1,...,N-1\}$). 

Let $d_{ij}$ denote a distance (directed graph edge) between the $i^{th}$ and $j^{th}$ cities. The tour length is given by
\begin{equation}
    H_{tour}= \sum_{i,j =0}^{N} d_{ij}\sum_{t=0}^{N-1} x_{i,t}x_{j,t+1}
\end{equation}
There exist two main constraints. 
First, one may travel to each city (vertex) only once. Consider the $i^{th}$ city. If one looks at a whole time cycle summing over all times one needs
\begin{equation}
    \sum_{t=0}^{N} x_{i,t}=1, \;\forall i\in \{0,1,...,N\}
\end{equation}
For instance, if the $i^{th}$ city was never visited, this sum would be $0$. If a city was visited more than once, this sum would be $> 1$. 
Second, one must travel to every city once (each time step contains only one city) producing the condition that
\begin{equation}
    \sum_{i=0}^{N} x_{i,t}=1, \forall t\in \{0,1,...,N\}
\end{equation}
Such a round trip tour is called a \textit{Hamilton cycle}.

Both these constraints must be added to the tour Hamiltonian $H_{tour}$ to give the TSP minimization function
\begin{equation}
\begin{array}{c}
H_{TSP} = H_{tour} + constraint \; terms \\\\
= \displaystyle\sum_{i,j =0}^{N} d_{ij}\sum_{t=0}^{N-1} x_{i,t}x_{j,t+1}  +\lambda \displaystyle\sum_{i=0}^{N}\left(\sum_{t=0}^{N}x_{i,t}-1 \right)^2 \\\\
    +\lambda \displaystyle\sum_{t=0}^{N}\left(\sum_{i=0}^{N}x_{i,t}-1 \right)^2
    \end{array}
\end{equation}
The ground state of this Hamiltonian is the solution of the TSP problem.
\subsubsection{Constraints}
For a problem of size $N$, there are $\binom{n}{2}$ constraints. The constraint matrices corresponding to these constraint terms must match the number of constraints \cite{SiddharthJain2021}. In the AQC formulation of the problem, enforcing hard constraints presents a challenge, as introducing high penalty terms often eliminates the feasibility of solutions. Therefore, it is crucial to establish bounded values for the Lagrange parameter to obtain the most optimal solutions. Additionally, chain strength plays a significant role in determining the feasibility and quality of returned solutions, ensuring that constraints are not violated.

Although D-Wave processors continue to improve, enabling the embedding of larger graphs, the majority of returned solutions remain suboptimal. Further research is necessary to identify the key parameters that significantly influence solution quality. Refining these parameters could enhance the likelihood of achieving optimal solutions with the highest number of valid samples.

\subsubsection{Normalization}
\begin{figure}[htbp!]
\centering
\includegraphics[width=1.0\linewidth]{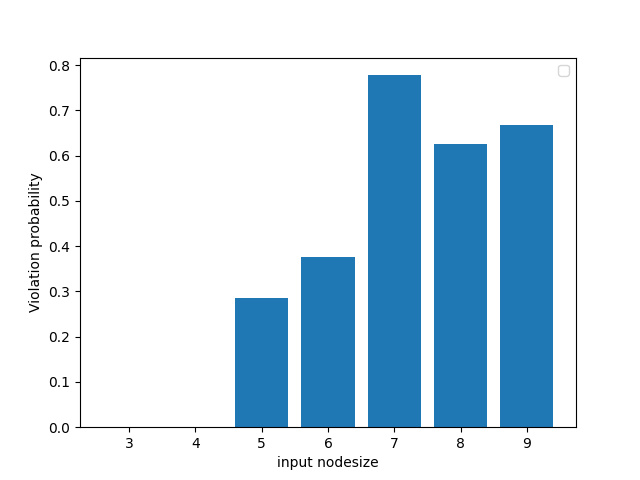}
\caption{Plot of Graph Size Vs. Violation Probability for random inputs.}
\label{"Constraint Violation Probability"}
\end{figure}

Since the AQC formulation is sensitive to variations in the Lagrange parameter, making it challenging to enforce constraints effectively, additional steps are necessary to carefully select an appropriate Lagrange parameter. A plot of constraint violation probability is shown in Figure \ref{"Constraint Violation Probability"} for test graphs with random weights. 

From recent efforts to enforce constraints rigorously before sampling energies and constructing solutions, we have observed that normalizing or scaling the matrix of weights significantly increases the likelihood of finding a feasible solution.

\begin{figure}[htbp!]
\centering
\includegraphics[width=1.0\linewidth]{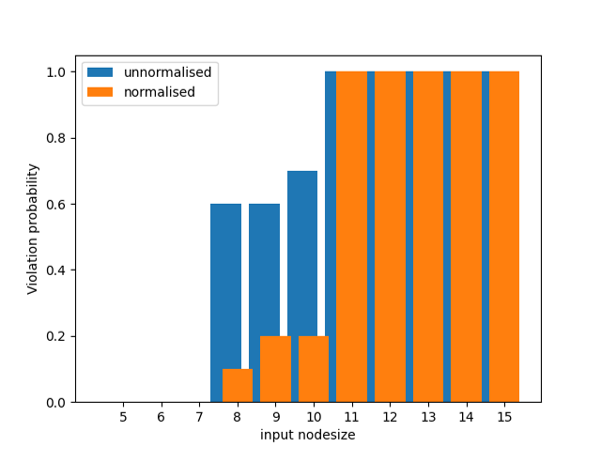}
\caption{Plot of Graph Size Vs. Violation Probability for selected test cases.}
\label{"Constraint Violation Probability with and without normalisation"}
\end{figure}

We conducted 10 test runs of the TSP code on each test graph for two cases: one without normalization and the other with normalization applied. As shown in Table \ref{norm table}, normalizing the weight matrix enhances the probability of obtaining a feasible solution. Figure \ref{"Constraint Violation Probability with and without normalisation"} compares the constraint violation probability for test graphs with random weights on the D-Wave Advantage6.1 machine, both with and without normalization.

\subsubsection{Hardware architecture}
The D-Wave platform provides quantum computing capabilities through its signature superconducting hardware, which leverages quantum tunneling across potential barriers to reach the ground state.

The D-Wave QPU consists primarily of nodes made from superconducting Josephson junctions that function as two-state qubits, forming a binary system. These qubits are coupled using RF-SQUIDs \cite{RF_SQUID}. At near-zero Kelvin temperatures, the system's possible energy states collapse to a singular value, corresponding to the lowest energy state or ground state. This principle, rooted in well-established thermodynamics, forms the fundamental operational mechanism of D-Wave hardware.

\begin{table}[htbp!]
\centering
\scalebox{0.7}{
\begin{tabular}{ |c|c|c| } 
\hline
Graph-size &
solutions without normalization & solutions
with normalization \\
\hline
8 & 4 & 9\\
9 & 4 & 8\\
10 & 3 & 8\\
\hline
\end{tabular}}
\caption{The table shows the number of runs vs the number of feasible solutions with and without normalization.}
\label{norm table}
\end{table}

\subsubsection{Solution for an 8 node TSP solved by the Dwave machines}
\begin{figure}
    \centering
    \begin{subfigure}
        \centering
        \includegraphics[scale=0.5]{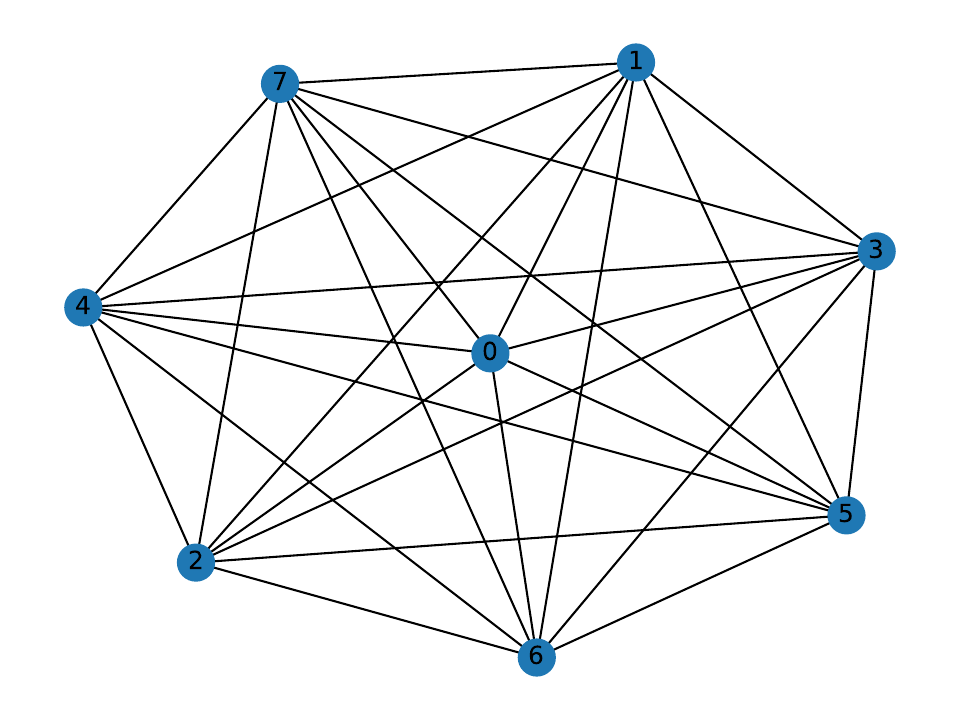}
        \caption{8 node input graph}
        \label{8 node input graph}
    \end{subfigure}
    \hfill
    \begin{subfigure}
        \centering
        \includegraphics[scale=0.5]{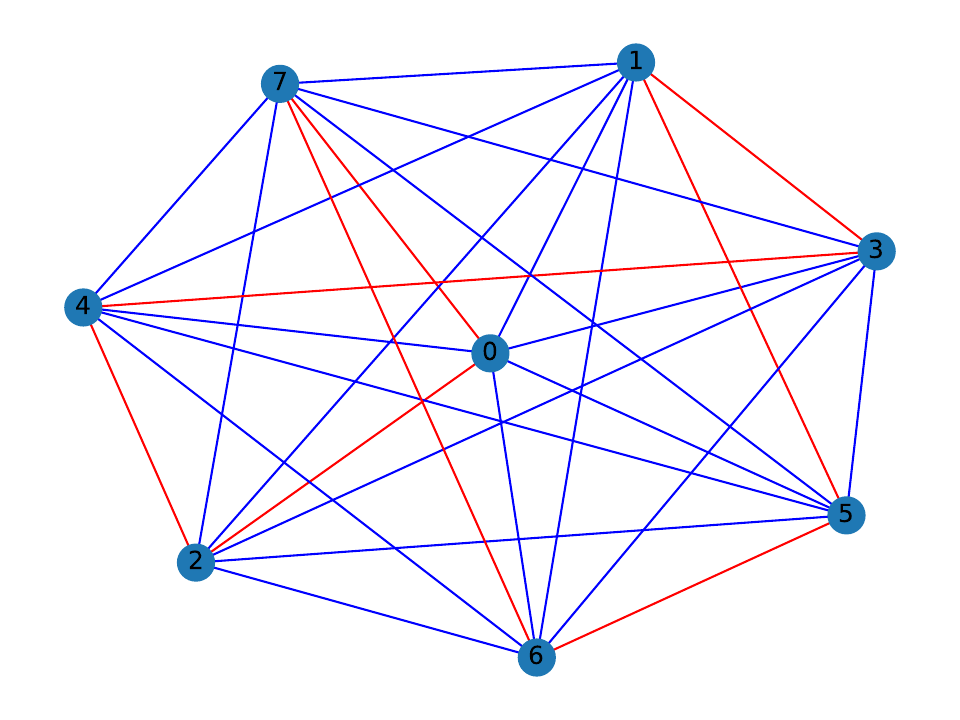}
        \caption{8 node output graph}
        \label{8 node output graph}
    \end{subfigure}
    \hfill
    \begin{subfigure}
        \centering
        \includegraphics[scale=0.5]{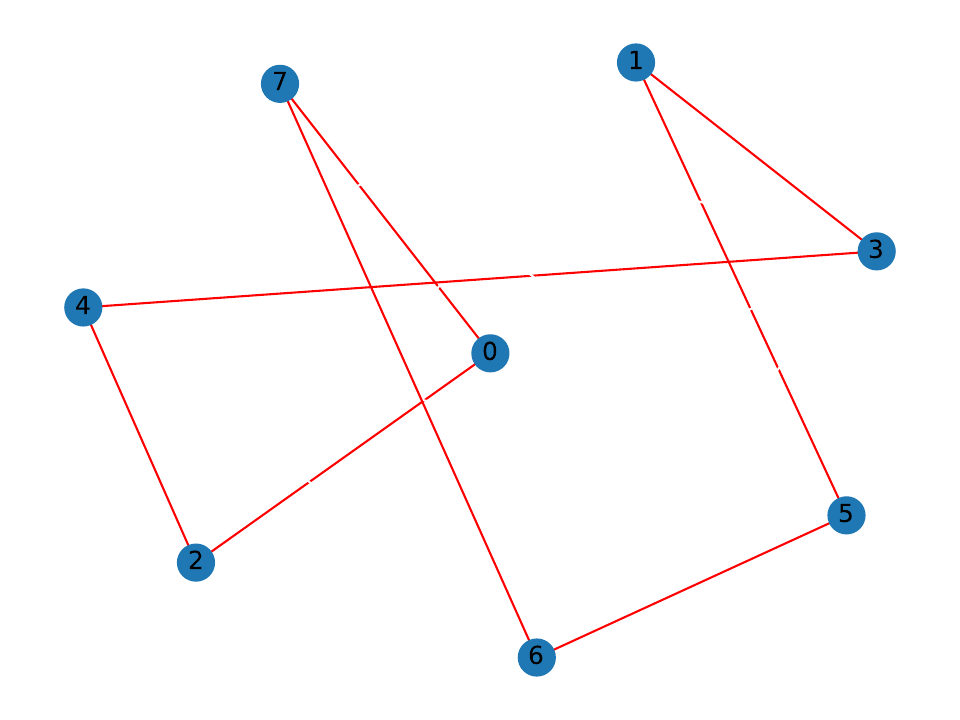}
        \caption{8 node output graph with the highlighted solution path}
        \label{8 node output graph white}
    \end{subfigure}
    \caption{The 8 node graph solved by D-wave quantum annealers. Plot (a) is the problem graph and plot (b) is the solution graph. Plot (c) is the highlighted solution path }
    \label{8 node graph solved by D-wave quantum annealers}
\end{figure}

An 8 node graph and its solution is shown in figure \ref{8 node graph solved by D-wave quantum annealers}

\subsection{Experimental Observations for Optical Ising computers}
The large-scale optical Ising machine represents a significant advance in photonic computing. This system utilizes a spatial light modulator (SLM), a device that imposes spatially varying modulation on a beam of light. In this configuration, the spin variables of the Ising problem are encoded through binary phase modulation of the light field \cite{Zhang}. Essentially, the phase of the light field corresponding to the position of the wave at a given point in space is modified in a binary manner to represent the spin variables.

A key challenge facing optical Ising machines is the difficulty in encoding the coupling coefficients $J_{ij}$ onto the SLM for a spatial optical Ising machine. One potential solution to this issue is the application of gauge transformations \cite{Gauge}, which provides an alternative approach to improve the efficiency and accuracy of encoding.

\subsubsection{Hamiltonian formulation on the QCi Dirac platforms}

QCi is currently offering hybrid quantum optimization machines that can solve discrete and quasi-continuous variables (Dirac-3) or QUBO/Ising problems (Dirac-1)
like D-wave quantum annealers. Dirac-3 has been shown to additionaly solve integer linear programming models and mixed integer linear programming models with a solution space spanning a range of integers without explicitly converting such a problem to binary. This allows us to use ILP formulations directly without any conversions to define the TSP to be solved on the QCI platforms. Two most popular ILP formulations that are being tested here are the Miller–Tucker–Zemlin \cite{CAMPUZANO_MTZ} formulation and the Dantzig–Fulkerson–Johnson formulation \cite{Applegate}.

\begin{equation} \text{min} \sum_{i=1}^n \sum_{j=1}^n c_{ij} x_{ij} \end{equation} where \( c_{ij} \) is the cost of traveling from city \( i \) to city \( j \), and \( x_{ij} \) is a binary variable indicating whether the path from \( i \) to \( j \) is included in the tour. 

\subsubsection{Subtour Elimination Constraints} Introduce variables \( u_i \) to represent the order in which cities are visited: \begin{equation} u_j - u_i \geq 1 - (n-1)(1 - x_{ij}) \quad \forall i, j \in \{2, \ldots, n\}, i \neq j \end{equation} These constraints ensure that no subtours are formed. 

\subsubsection{Additional Constraints} Each city must be visited exactly once: \begin{equation} \sum_{j=1}^n x_{ij} = 1 \quad \forall i \end{equation} \begin{equation} \sum_{i=1}^n x_{ij} = 1 \quad \forall j \end{equation} The order variables must be within a valid range: 

\begin{equation} 2 \leq u_i \leq n \quad \forall i \in \{2, \ldots, n\} \end{equation}

\subsubsection{Hardware architecture}
Entropy  quantum  computing was recently introduced as an efficient optimization technique that leverages the principles of entropy  minimization to guide quantum state evolution toward optimal solutions  of a Hamiltonian \cite{nguyen_entropy_2024}. In a photonic realization, qudits are encoded as a superposition of photon numbers in time bins that evolve in an optical fiber loop which embeds a desired Hamiltonian as a dissipative operator.  This effectively emulates imaginary time  evolution, in which the propagation of higher-energy eigenstates is  subject to dissipation and decoherence while the lower-energy  eigenstates are promoted in this evolution. 

The current generation of the Entropy Quantum Computing (EQC) hardware  at QCi, named Dirac-3, is a hybrid system that harnesses the high-speed parallel processing capabilities of photonics and the quantum nature of light with the precise control and programmability of electronic circuits. By leveraging photonic systems for entropy-driven state evolution—enabled by their natural ability to manipulate complex optical fields—and utilizing electronics for state initialization, feedback,  and fine-tuning, hybrid entropy computing achieves  efficient optimization.  Such systems can explore large solution spaces rapidly through photonic modes propagation while employing electronic components to refine solutions iteratively.  This synergy offers a pathway for solving challenging optimization problems in areas such as machine learning, signal processing, and network design, with the potential for high speed, low power consumption, and scalability.

The EQC hardware Dirac-1 targets optimization of a polynomial objective function $E$ in the following form:
\begin{equation}
    E = \sum_{i} C_i x_i +
        \sum_{i,j} J_{ij} x_i x_j .
\end{equation}
where, $x_i\in \{0,1\}$ are optimization variables, $C_i$ are real-valued coefficients of linear terms, and $J_{ij}$ represent two-body interaction coefficients that are real numbers subject to the tensor $J$ being symmetric under all permutations of the indices. Governed by the physical mechanism of operation of the EQC hardware, $x_i$ represents photon numbers in the $i$’th time bin degrees of freedom. 

A major advantage of EQC compared to other quantum annealers is its flexible variable encoding capability \cite{nguyen_entropy_2024}. While, in the form discussed above, EQC allows for solving continuous-variable optimization problems, by allocating multiple photon time bins to one variable, one can effectively encode binary or integer variables for solving combinatorial optimization problems in the forms of binary, integer or even mixed-integer problems.


Researchers have shown that by doing this, the propagation of light can be tailored or adjusted to minimize an Ising Hamiltonian. Using Min-Max normalization to ensure a successful Hamilton cycle, the photonic Ising machine Can solve the TSP for a problem upto 18 nodes for any linear constraints of the type $Ax = b$. 

With the advantage of all-to-all connectivity 
no explicit embedding step is required as opposed to the D-wave platform \cite{gilbert2024benchmarkingquantumannealersnearoptimal}, the QCI Dirac machines are able to solve TSP instances of sizes beyond 12 nodes.

\subsubsection{Solution for an 18 node TSP}
\begin{figure}
    \centering
    \begin{subfigure}
        \centering
        \includegraphics[scale=0.5]{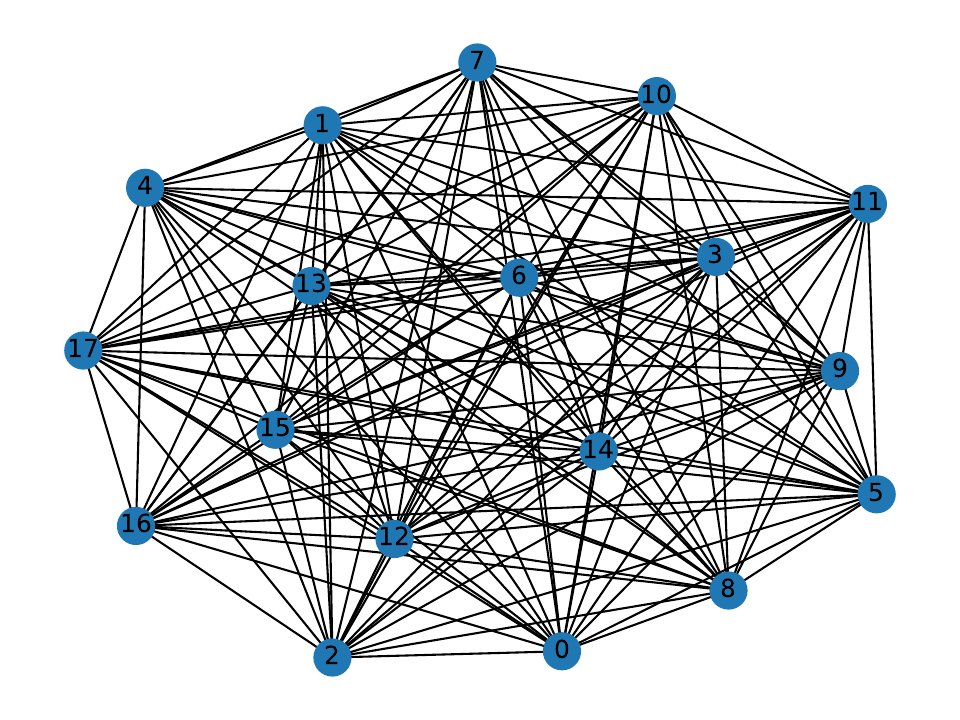}
        \caption{18 node input graph}
        \label{18 node input graph}
    \end{subfigure}
    \hfill
    \begin{subfigure}
        \centering
        \includegraphics[scale=0.5]{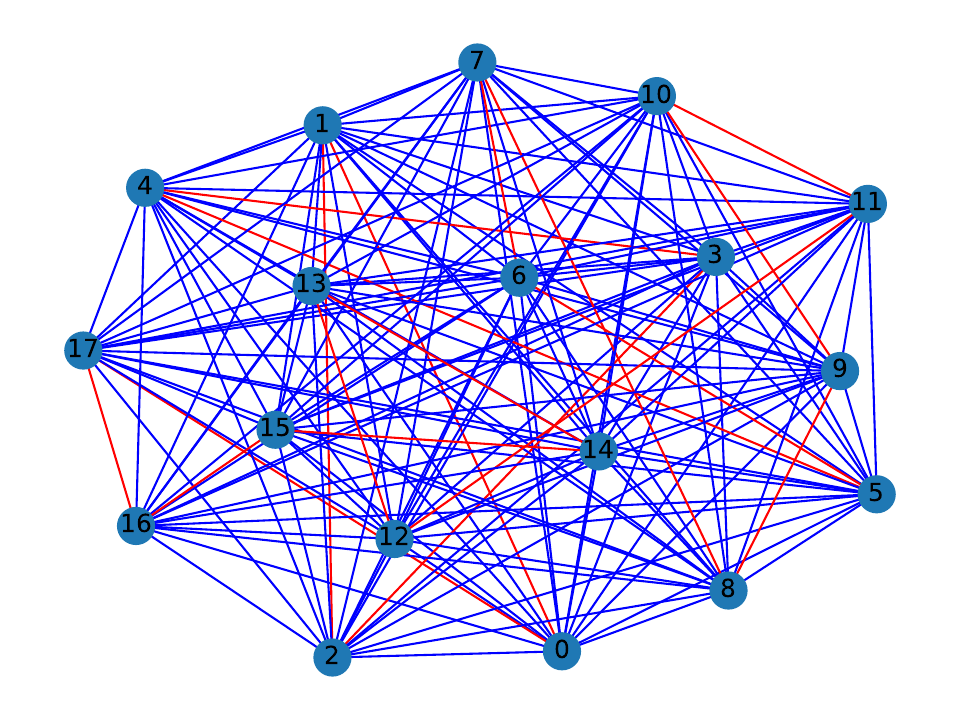}
        \caption{18 node output graph}
        \label{18 node output graph}
    \end{subfigure}
    \hfill
    \begin{subfigure}
        \centering
        \includegraphics[scale=0.5]{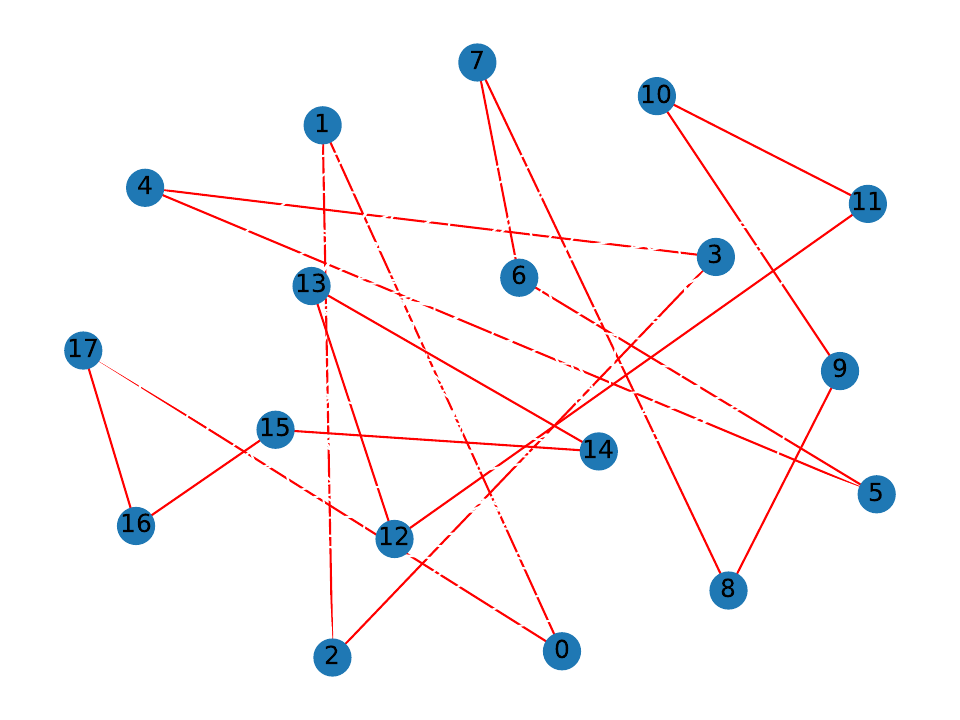}
        \caption{18 node output graph with the highlighted solution path}
        \label{18 node output graph white}
    \end{subfigure}
    \caption{The 18 node graph solved by QCI Dirac quantum annealers. Plot (a) is the problem graph and plot (b) is the solution graph, Plot (c) is the solution path in the graph }
    \label{18 node graph QCI Dirac machines}
\end{figure}

 The maximum size of the TSP that can be solved reliably without violating the constraints is currently 18 nodes. Figures \ref{18 node graph QCI Dirac machines} is a demonstration of the 
 Dirac-1 
 machine solving a TSP instance of 18 nodes.

\section{Results}
We can clearly see that the Traveling Salesman Problem (TSP) is well-suited for direct Ising-type processors, as the cost function is inherently quadratic, contrasting with the quantum algorithms that utilize universal gate-based quantum processors. With ongoing advancements in the design and functionality of Ising-type quantum machines, we have not only demonstrated proofs of concept for small test cases but also addressed the scalability challenges. These improvements bring us closer to developing robust practical applications for solving real-world optimization problems.

Although our results confirm that Ising-based processors can effectively solve the TSP \cite{si2023energyefficientsuperparamagneticisingmachine}, it remains crucial to verify whether the returned solutions are optimal. A common challenge with current Ising-based machines is their tendency to become trapped in local minima rather than converging to the global minimum \cite{suboptising}. A thorough assessment is required to compare the obtained solutions with classical counterparts to ensure that the identified Hamiltonian cycle is not only valid but also represents the true minimum weight configuration.

Verification against classical solutions reveals that Ising processors do not always return the ground state solution and often produce suboptimal results. However, for smaller problem instances, Ising machines exhibit higher accuracy in finding optimal solutions. To enhance the reliability of the results, multiple solution samples should be collected and statistically evaluated. The probability of obtaining the true ground state solution improves with an increasing number of sampled runs.

\begin{figure}[htbp!]
\centering
\includegraphics[width=1.0\linewidth]{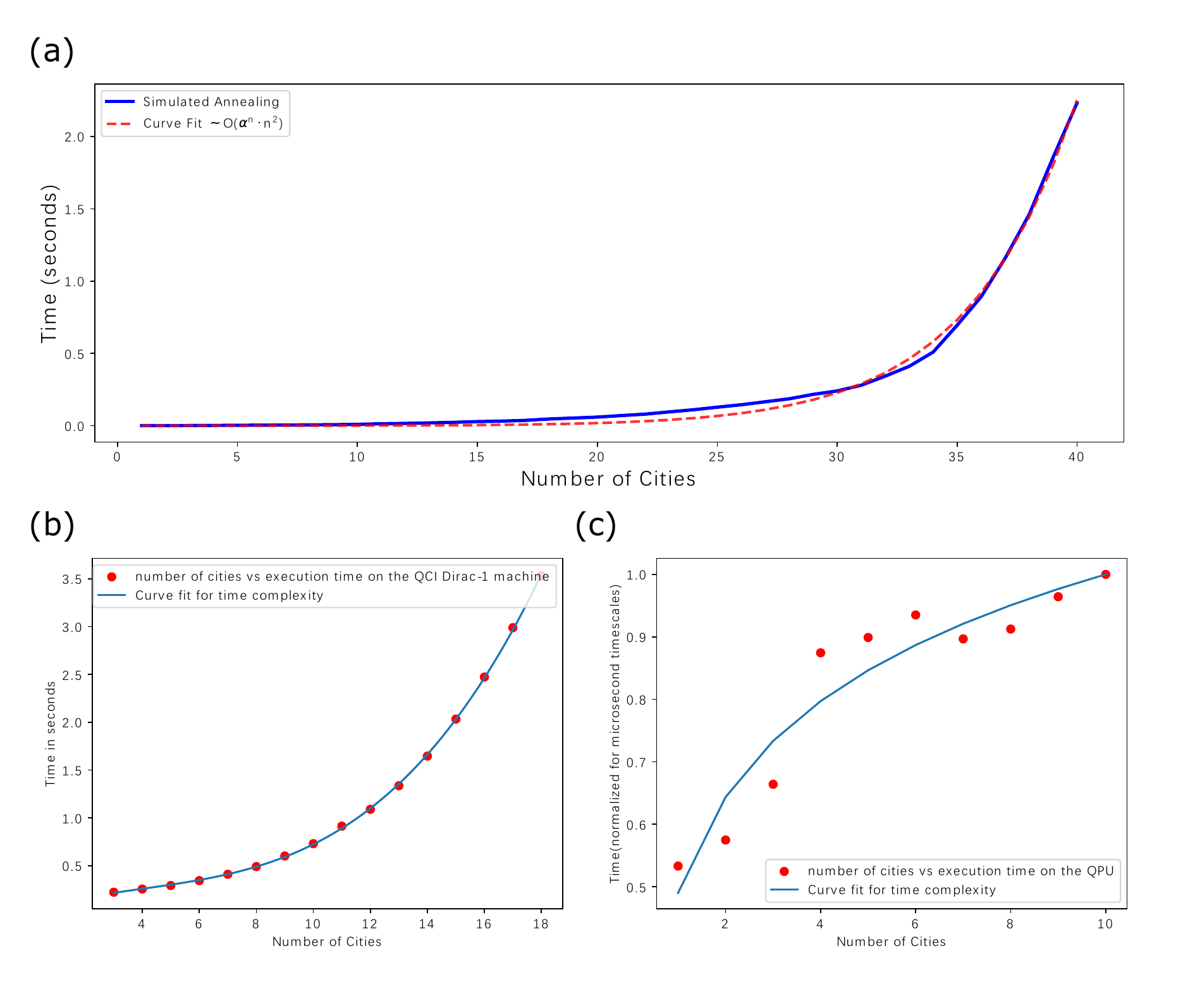}
\caption{This is a plot of the algorithm run time for (a) Simulated annealing for TSP instances of sizes upto 40 (b) QCI-Dirac1 for TSP instances of sizes upto 18 (c) D-wave Advantage4.1 for TSP instances of sizes upto 10}
\label{"Time complexities"}
\end{figure}

\begin{table}[htbp!]
    \centering
    \begin{tabular}{|c|c|}
    \hline
         TSP Algorithm & Time Complexity \\
    \hline
        Brute Force & $O(n!n)$\\ 
    \hline
        Christofides algorithm & $O(n^3)$\\
    \hline
        Christofides with Serdyukov approximation & $O(n^3 \log n)$\\
    \hline
        Nearest Neighbour algorithm & $O(n^2)$\\
    \hline
        Dynamic Programming & $O(2^n*n^2)$\\
    \hline
        Simulated Annealing & $O(1.62^n*n^2)$\\
    \hline
        Ambiani's et. al.& $O(1.728^n)$\\
    \hline
        QUBO on Dwave (best case) & $O(2\log n))$\\
    \hline
        QUBO on Dirac-1 & $O(n^{3/2}))$\\
    \hline
    \end{tabular}
    \caption{time complexity table}
    \label{timecomplexity table}
\end{table}

Adiabatic Quantum Computers (AQC) are indeed capable of solving Quadratic Mixed Integer Programming (MLIP) problems. This is because these problems can be formulated as Quadratic Unconstrained Binary Optimization (QUBO) problems, which are well-suited for AQCs.

For example, in the field of machine learning, AQCs have been utilized to train models such as Support Vector Machines (SVMs) \cite{date2024adiabaticquantumsupportvector}. The training process, which involves solving an optimization problem, can be reformulated as a QUBO problem and effectively tackled using an AQC \cite{Date_2021}. In particular, the time complexity of this quantum approach has been demonstrated to be an order of magnitude better than that of classical methods \cite{Wangultra}.

Furthermore, empirical studies have shown that the quantum approach achieves a 3.5–4.5 times speedup over classical methods when applied to datasets containing millions of features. This indicates that AQCs have the potential to provide substantial advantages in solving Quadratic MLIP problems, particularly for large-scale datasets \cite{Sannia_2023} \cite{wang2022quantuminspired}.

Additionally, researchers have demonstrated a quantum speedup for the Traveling Salesman Problem (TSP) on bounded-degree graphs. Specifically, the improvement in computational speed has been shown to be quadratic when the degree of each vertex is at most three \cite{Moylett_2017}.

\section{Conclusion}
The time complexities for Ising-type processors are rough estimates based on the curve fitting method, as shown in Figure \ref{"Time complexities"}. A more rigorous benchmarking approach is required for improved accuracy in determining time complexity, particularly as constraint violation probability increases with problem size.

Table \ref{timecomplexity table} presents a comparison between different approaches. It is evident that Ising-type processors provide a definitive speedup for solving specific classes of problems compared to traditional classical computing architectures.

We have demonstrated TSP instances being solved using novel optical, optoelectronic, and quantum hardware and conducted benchmarks to assess the performance limits of currently available commercial machines based on these architectures. NISQ-era universal gate-based quantum computers, which utilize QAOA and QPE, currently struggle to solve TSP instances exceeding six nodes and rely on classical optimization to determine optimal solutions in the case of QAOA. Quantum annealers have successfully solved TSP instances of up to ten nodes with near-optimal solutions. Meanwhile, optoelectronic Ising machines have demonstrated the ability to handle up to eighteen nodes reliably, although solutions tend to be suboptimal for larger instances. As these technologies continue to evolve, we anticipate that increasingly complex TSP instances will become solvable with greater efficiency and accuracy. 

\section*{Acknowledgement}
This work was supported in part by the ACC-New Jersey under Contract No. W15QKN-24-C-0004.

\end{document}